\documentstyle[epsfig]{article}
\newcommand{\bbox}[1]{\mbox{\boldmath ${#1}$}}

\begin{document}

\title{A complete set of total cross sections for imaginary parts of $nd$ forward scattering amplitudes and three-nucleon force effects}
\author{S. Ishikawa and M. Tanifuji\\
Department of Physics, Hosei University,\\
Fujimi 2-17-1, Chiyoda, Tokyo 102-8160, Japan\\
and\\
Y. Iseri\\
Department of Physics, Chiba-Keizai College, \\
Todoroki-cho 4-3-30, Inage, Chiba 263-0021, Japan}

\maketitle
\begin{abstract}
In the neutron-deuteron scattering, four total cross sections are shown to form a complete set for the determination of the imaginary parts of the forward amplitudes because of the optical theorem.
The amplitudes are decomposed into scalar and tensor ones in the spin space.
Contributions of three-nucleon forces (3NF) to these amplitudes are studied, by the Faddeev calculation, which predicts significant tensor effects of the 3NF.
\end{abstract}


In recent years, three-nucleon systems have attracted a considerable attention as important sources of information on nuclear interactions, because of possible participation of three-nucleon forces (3NF) in addition to two-nucleon ones.
Various versions of the two-nucleon force (2NF) have been examined as the input of three-nucleon (3N) calculations and have been found to be deficient in reproducing some empirical observables\ \cite{Gl96}.
Improvements of the calculations have been achieved by introducing the $2\pi$ exchange 3NF on the three-nucleon binding energy\ \cite{Sa86} and the minimum of the proton-deuteron ($pd$) elastic scattering cross section between 50 and 200 MeV\ \cite{Wi98}.
However, the prescription is not always effective for observables in polarization phenomena.
In fact, the 3NF cannot explain the proton vector analyzing power between 65 and 200 MeV\cite{St99} of $\vec{p}d$ scattering, and the deuteron tensor analyzing powers at 270 MeV of $\vec{d}p$ scattering\ \cite{Sa00}. 
Also, discrepancies of the nucleon and deuteron vector analyzing powers between the calculated and the measured in low energy neutron-deuteron ($nd$) scattering still remain to be unsolved even with the 3NF\ \cite{Wi94,Is99}.

These indicate that the comprehensive understanding about a role of the nuclear interactions, particularly that of their spin dependence, on the 3N spin observables has not yet been obtained.
In general, the spin dependence of the nuclear interactions should be reflected in the spin observables through the spin structure of scattering amplitudes.
Paying attention to the spin structure, we will consider imaginary parts of forward amplitudes of the $nd$ elastic scattering, which are described in terms of total cross sections by the optical theorem, to criticize the nuclear interactions in a different aspect.

Total cross sections have varieties due to the combination of the spin orientation of projectiles and that of targets.
Denoting  spin density matrices of the neutron and the deuteron in the initial state by $\rho^{(n)}$ and $\rho^{(d)}$\ \cite{Oh72}, respectively, 
and the scattering matrix at the forward angle by $\bbox{M}_{\theta=0}$, 
the corresponding total cross section $\sigma^{tot}$ is given by the optical theorem as
\begin{equation}
\sigma^{tot}=\alpha {\rm Im}\left\{ {\rm Tr}(\rho^{(n)} \rho^{(d)} \bbox{M}_{\theta=0}) \right\},
\label{eq:opt-thrm}
\end{equation}
where 
\begin{equation}
\alpha= \frac{4\pi}k,
\end{equation}
with $k$ being the magnitude of the $nd$ relative momentum $\bbox{k}$.
When such kinds of total cross sections are measured sufficiently, the determination is unambiguous and the imaginary parts obtained will become a reliable scale for the criticism of the calculated scattering amplitudes and then the nuclear interactions employed.  

In the following, we will give the independent forward amplitudes of the $nd$ elastic scattering and derive in a systematic way the total cross sections of the complete set for the imaginary parts of the amplitudes which include the previous works\ \cite{Ab98,Wi99,Wi999}. 
Then, we will investigate the spin structure of the $nd$ scattering amplitude by decomposing the amplitude by the spin space tensors.
Finally, effects of 3NF on these amplitudes are examined by the Faddeev calculations.

In the $nd$ scattering, we have four non-vanishing independent forward amplitudes \cite{Re95}.
Designating elements of $\bbox{M}$ by the $z$ components of spins of the related particles $\nu$ as 
$\langle \nu_n^\prime, \nu_d^\prime |\bbox{M}|\nu_n, \nu_d \rangle_{\theta=0}$, these are,
\begin{eqnarray}
M_1 &=& \langle \frac12, 1|\bbox{M}| \frac12, 1 \rangle_{\theta=0},
\nonumber\\
M_2 &=& \langle -\frac12, 1 |\bbox{M}| -\frac12, 1 \rangle_{\theta=0},
\nonumber\\ 
M_3 &=& \langle -\frac12, 1 |\bbox{M}| \frac12, 0 \rangle_{\theta=0}
  = \langle \frac12, 0 |\bbox{M}| -\frac12, 1 \rangle_{\theta=0},
\nonumber\\
M_4 &=& \langle \frac12, 0 |\bbox{M}| \frac12, 0 \rangle_{\theta=0}.
\label{eq1}
\end{eqnarray}
Then, in principle, four kinds of independent total cross sections determine all of the imaginary parts of the forward scattering amplitudes due to the optical theorem.  

One of such total cross sections should be that for the unpolarized neutrons and deuterons, $\sigma_0^{tot}$, where spin density matrices are given by
\begin{equation}
\rho^{(n)}=\frac12 I^{(n)}, \quad \rho^{(d)}=\frac13 I^{(d)},
\end{equation}
where $I^{(n)}$ and $I^{(d)}$ are unit matrices.
We get from Eq.\ (\ref{eq:opt-thrm}),
\begin{equation}
\sigma_0^{tot}=\frac{\alpha}3 {\rm Im}(M_1+M_2+M_4).
\label{eq7}
\end{equation}

Next, we will choose the cross sections for vector-polarized neutrons and deuterons.
Recently two kinds of such cross sections have been theoretically investigated, focusing attention to cross section asymmetries, for which noticeable contributions of the 3NF have been predicted \cite{Wi999}.
When the neutron and the deuteron are vector-polarized in the same direction along the $z$ axis with polarizations $p_z^{(n)}$ and $p_z^{(d)}$, the corresponding cross section $\sigma_L^{tot}(p_z^{(n)}, p_z^{(d)})$ is obtained by Eq.\ (\ref{eq:opt-thrm}) with spin density matrices,
\begin{equation}
\rho^{(n)}= \frac12 p_z^{(n)} \sigma_z,
  \quad \rho^{(d)}=\frac12 p_z^{(d)} P_z,
\label{eq:rhoz}
\end{equation}
where
\begin{equation}
P_z = \left(
\begin{array}{ccc}
    1 & 0 & 0 \\
    0 & 0 & 0 \\
    0 & 0 & -1
\end{array}
\right).
\label{eq:Pz}
\end{equation}
On the other hand, when the neutron and the deuteron are vector-polarized in the same direction but the direction is perpendicular to the $z$ axis, one can choose the $y$ axis in the polarization direction with polarizations $p_y^{(n)}$ and $p_y^{(d)}$. 
Then the corresponding cross section $\sigma_T^{tot}(p_y^{(n)}, p_y^{(d)})$ is obtained by Eq.\ (\ref{eq:opt-thrm}) with spin density matrices,
\begin{equation}
\rho^{(n)}= \frac12 p_y^{(n)} \sigma_y,
 \quad \rho^{(d)}= \frac12 p_y^{(d)} P_y,
\label{eq:rhoy}
\end{equation}
where
\begin{equation}
P_y= \frac1{\sqrt2} \left(
     \begin{array}{ccc}
     0& -i& 0 \\
     i& 0& -i \\
     0& i& 0 \\
     \end{array}
     \right).
\label{eq:Py}
\end{equation}

The cross section asymmetries in Ref.\ \cite{Wi999} are defined as the cross section difference provided by the change of the deuteron spin direction to the opposite one as
\begin{eqnarray}
\Delta \sigma_L &=& \sigma_L^{tot}(+1, -1) - \sigma_L^{tot}(+1, +1) = -2 \sigma_L^{tot}(+1, +1)
\nonumber \\
\Delta \sigma_T &=& \sigma_T^{tot}(+1, -1) - \sigma_T^{tot}(+1, +1) = -2 \sigma_T^{tot}(+1, +1).
\nonumber \\
\label{eq:deltas}
\end{eqnarray}
Using Eqs.\ (\ref{eq:opt-thrm}) and (\ref{eq:deltas}) with Eqs.\ (\ref{eq:rhoz}), (\ref{eq:Pz}), (\ref{eq:rhoy}), (\ref{eq:Py}), we obtain
\begin{equation}
\Delta\sigma_L = - \alpha {\rm Im}(M_1 -M_2),
\label{eq15}
\end{equation}
and
\begin{equation}
\Delta\sigma_T = - \alpha \sqrt2 {\rm Im}(M_3).
\label{eq19}
\end{equation} 

As the last one, we  consider the total cross section for the scattering of the unpolarized neutron by the tensor polarized deuteron.
For the unpolarized neutron and the $t_{20}$ tensor polarized deuteron along the $z$ axis, we have
\begin{equation}
\rho^{(n)}=\frac12 I^{(n)},
  \quad \rho^{(d)}=\frac{\sqrt2}{6} t_{20} P_{zz},
\label{eq:rhozz}
\end{equation}
where
\begin{equation}
P_{zz} = \left(
\begin{array}{ccc}
1 & 0 & 0 \\
0 & -2 & 0 \\
0 & 0 & 1
\end{array}
\right).
\label{eq:Pzz}
\end{equation} 
We get the total cross section $\sigma_{20}^{tot}$ for $t_{20}=1$
\begin{equation}
\sigma_{20}^{tot}=\frac{\alpha}{3\sqrt{2}} {\rm Im}(M_1 +M_2 -2M_4).
\label{eq11}
\end{equation}

Solving Eqs.\ (\ref{eq7}),  (\ref{eq15}),  (\ref{eq19}), and (\ref{eq11}),
one gets the imaginary parts of the forward scattering amplitudes $M_1 \sim M_4$ in terms of the total cross sections,
\begin{eqnarray}
\alpha {\rm Im}(M_1)&=&\sigma_0^{tot} +\frac1{\sqrt2}\sigma_{20}^{tot} - \frac12 \Delta\sigma_L,
\nonumber\\
\alpha {\rm Im}(M_2)&=&\sigma_0^{tot} +\frac1{\sqrt2}\sigma_{20}^{tot} + \frac12 \Delta\sigma_L,
\nonumber\\
\alpha {\rm Im}(M_3)&=&- \frac1{\sqrt2} \Delta\sigma_T,
\nonumber\\
\alpha {\rm Im}(M_4)&=&\sigma_0^{tot} -{\sqrt2} \sigma_{20}^{tot}.
\label{eq22}
\end{eqnarray}
By these equations, one can determine the imaginary parts of the scattering amplitudes $M_1 \sim M_4$ unambiguously when $\sigma_0^{tot}$, $\Delta\sigma_L$, $\Delta\sigma_T$, and $\sigma_{20}^{tot}$ are measured. 
The present choice of the set of the independent total cross sections is not unique.  
However, different choices will produce the information of the scattering amplitudes equivalent to the present one, since the independent amplitudes are restricted to four ones.

In order to get deeper insights into spin dependence of the interactions, we decompose the scattering matrix $\bbox{M}$ by spin space tensors of rank $K$ with $z$ component $\kappa$, $\bbox{S}_{K\kappa}$,
\begin{equation}
\bbox{M} = \sum_{K \kappa} (-)^{\kappa} \bbox{S}_{K, -\kappa} 
  \bbox{R}_{K\kappa},
\label{eq231}
\end{equation}
where $\bbox{R}_{K\kappa}$ is the counterpart, the tensor in the coordinate space. 
The matrix element of $\bbox{M}$ for a reaction A(a,b)B is given in terms of invariant amplitudes\ \cite{Ta98},
\begin{eqnarray}
&& \langle \nu_b, \nu_B; \bbox{k}_f | \bbox{M} |\nu_a, \nu_A; \bbox{k}_i \rangle
\nonumber\\
  &=& \sum_{s_i s_f K} (s_a s_A \nu_a \nu_A|s_i \nu_i) (s_b s_B \nu_b \nu_B|s_f \nu_f) 
\nonumber\\
 & & \times (s_i s_f \nu_i, -\nu_f|K \kappa) (-)^{s_f-\nu_f} 
 \sum_{\ell_i=\bar{K}-K}^K 
\nonumber\\ 
 & & \times 
  [C_{\ell_i}(\hat{k}_i) \otimes C_{\ell_f=\bar{K}-l_i}(\hat{k}_f)]_{\kappa}^K 
  F(s_i s_f K \ell_i),
\label{eq:inv-amp}
\end{eqnarray}
where $\bbox{k}_i$ ($\bbox{k}_f$) is the relative momentum in the initial (final) state, 
$s$ ($\nu$) denotes the spin ($z$ component) 
and $\bar{K}$ is $K$ for $K=$ even and $K+1$ for $K=$ odd\ \cite{Ta68}.
The quantum number $s_i$ ($s_f$) is the channel spin for initial (final) state.
The quantity $\hat{k}_i$ ($\hat{k}_f$) is the solid angle of $\bbox{k}_i$ ($\bbox{k}_f$) and $C_{\ell m}(\hat{k})$ is related to the spherical harmonics  $Y_{\ell m}(\hat{k})$ as usual\ \cite{de63}.
In Eq.\ (\ref{eq:inv-amp}), the geometrical parts of the matrix element of the tensors are given by the Clebsch-Gordan coefficients and $[C_{\ell_i}(\hat{k}_i) \otimes C_{\ell_f}(\hat{k}_f)]_{\kappa}^K$, and their physical parts are included in $F(s_i s_f K \ell_i)$, the invariant amplitude, which is a function of the scattering angle and the center-of-mass energy, although omitted for simplicity.
The amplitude $F(s_i s_f K \ell_i)$ describes the scattering by the interaction designated by the tensor rank $K$ in the spin space:
for example $F(s_i s_f K=0\ \ell_i)$ represents the scattering by scalar interactions, that is, central interactions in the sense of effective interactions, which include any higher order of the interactions as long as it forms the scalar in the spin space. 
More details are given in Refs.\ \cite{Ta98,Ta68}.

In the present case, $\bbox{k}_i = \bbox{k}_f = \bbox{k}$ and $z \parallel \bbox{k}$,  we have non-vanishing two scalar amplitudes, $U_1$ and $U_3$, and two tensor ones, $T_1$ and $T_3$ defined as
\begin{eqnarray}
U_{2s} &=& F(s s 0 0)
\nonumber\\ 
T_{2s} &=& F(\frac32 s 2 0) +\sqrt{\frac23}F(\frac32 s 2 1) + F(\frac32 s 22),
\end{eqnarray}
where $s = 1/2$ (the doublet state) and $3/2$ (the quartet state).
From Eqs.\ (\ref{eq22}) and (\ref{eq:inv-amp}), we obtain
\begin{eqnarray}
\alpha {\rm Im}(U_1) &=& \sqrt2 \sigma_0^{tot} + \frac{\sqrt2}{3}(\Delta\sigma_L + 2 \Delta\sigma_T), 
\nonumber\\
\alpha {\rm Im}(U_3) &=& 2 \sigma_0^{tot} - \frac13(\Delta\sigma_L + 2 \Delta\sigma_T),
\nonumber\\
\alpha {\rm Im}(T_1) &=& -\sqrt2\sigma_{20}^{tot} - \frac13 (\Delta\sigma_L - \Delta\sigma_T),
\nonumber\\
\alpha {\rm Im}(T_3) &=& \sqrt2\sigma_{20}^{tot} - \frac23 (\Delta\sigma_L - \Delta\sigma_T).
\label{eq25-3}
\end{eqnarray}

The amplitudes $U_1 \sim T_3$ are general.
Then, to connect them with realistic interactions, we will choose an explicit form for $\bbox{M}_{\theta=0}$ as, 
\begin{eqnarray}
\bbox{M}_{\theta=0} &=&
S_0 + S_\sigma \left( \bbox{s}_n \cdot \bbox{s}_d \right)
 + W_D  [\bbox{s}_d \otimes \bbox{s}_d]^2_0
\nonumber \\
&& + W_T  [\bbox{s}_n \otimes \bbox{s}_d]^2_0,
\label{eq:M0}
\end{eqnarray}
where $\bbox{s}_n$ and $\bbox{s}_d$ are the spin operators of neutron and deuteron.
Here, $S_0$ and $S_\sigma$ are the space parts of scalar amplitudes, and $W_D$ and $W_T$ are for tensor ones.
These amplitudes are related to the invariant amplitudes, $U_1$, $U_3$, $T_1$, and $T_3$, as
\begin{eqnarray}
S_0 &=& \frac13 \left( U_3+ \frac1{\sqrt2} U_1 \right),
\nonumber \\
S_\sigma &=& \frac13 \left( U_3 - \sqrt{2} U_1 \right),
\nonumber \\
W_D &=& \frac12 \left( T_3 - 2 T_1 \right),
\nonumber \\
W_T &=& T_3 + T_1,
\end{eqnarray}
which lead to the following equations:
\begin{eqnarray}
\alpha {\rm Im}(S_0) &=& \sigma_0^{tot}, 
\nonumber\\
\alpha {\rm Im}(S_\sigma) &=& -\frac13 (\Delta\sigma_L + 2 \Delta\sigma_T),
\nonumber\\
\alpha {\rm Im}(W_D) &=& \frac{3}{\sqrt2}\sigma_{20}^{tot},
\nonumber\\
\alpha {\rm Im}(W_T) &=& - (\Delta\sigma_L - \Delta\sigma_T).
\label{eq28}
\end{eqnarray}

If we consider a simple folding potential between the neutron and the deuteron neglecting antisymmetrization and other reaction mechanism, the relation between $S_0 \sim W_T$ and nuclear force component turns to be rather straightforward.
Let us assume the nuclear force between nucleons $i$ and $j$ to consist of spin-independent, spin-spin, and tensor forces as,
\begin{equation}
V_{i,j} = V_0(i,j) + V_\sigma(i,j) (\sigma_i\cdot\sigma_j) + V_T(i,j) S_T(i,j).
\end{equation}
In the first order approximation, it is easily shown that $S_0$ and $S_\sigma$ are provided by the scalar interactions, $V_0$ and $V_\sigma$, respectively, with the $S$-state component of the deuteron internal wave function, $W_D$ is derived from $V_0$ with the $D$-state component, and $W_T$ from $V_T$ with the $S$-state component.
Therefore, the measurements of $\sigma_0^{tot}$, $\Delta\sigma_L$, $\Delta\sigma_T$, and $\sigma_{20}^{tot}$ would provide rather pure information on the respective interactions.
From Eq.\ (\ref{eq28}), one will see that $\sigma_0^{tot}$ is given only by the imaginary part of the spin-independent scalar amplitudes, and $\sigma_{20}^{tot}$ by that of the tensor one whose origin can be considered as the deuteron $D$-state.
On the other hand, the cross section asymmetries $\Delta\sigma_L$ and $\Delta\sigma_T$,  contain information on the imaginary part of the spin-dependent scalar amplitude and that of the intrinsic tensor amplitude.
The quantity $\Delta\sigma_L - \Delta\sigma_T$ provides rather direct information of the  nuclear tensor interaction.

We will predict the total cross sections for the complete set at low incident energies by the Faddeev calculation, in which the 2NF is fixed to the Argonne V$_{18}$ potential (AV18)\ \cite{Wi95} while the 3NF is the $2\pi$ exchange Brazil model (BR-3NF)\ \cite{Ro86} with the cut-off parameter adjusted so as to reproduce the empirical triton binding energy.
Due to the $2\pi$ exchange mechanism, the BR-3NF is expected to contribute to not only scalar nuclear forces but also tensor ones in the spin space.
In order to demonstrate the role of the tensor forces, a fictitious Gaussian 3NF which is the spin independent force (GS-3NF),
\begin{equation}
V_G=V_0^G \exp\{-(\frac{r_{21}}{r_G})^2 -(\frac{r_{31}}{r_G})^2 \} +(c.p.),
\label{eq30}
\end{equation}
is examined with $r_G=1.0$ fm and $V_0^G=-45$ MeV, which are fixed so as to reproduce the empirical triton binding energy.

Numerical calculations of the low energy $nd$ scattering are performed by solving the 3N Faddeev equation in coordinate space \cite{Sa86,Is87,Is95}.
In the present calculation, 3N partial wave states for which 2NF and 3NF act, are restricted to those with total NN angular momenta $j \le 2$.
The total 3N angular momentum ($J$) is truncated at $J=19/2$.

\begin{figure}
\begin{center}
\epsfig{file=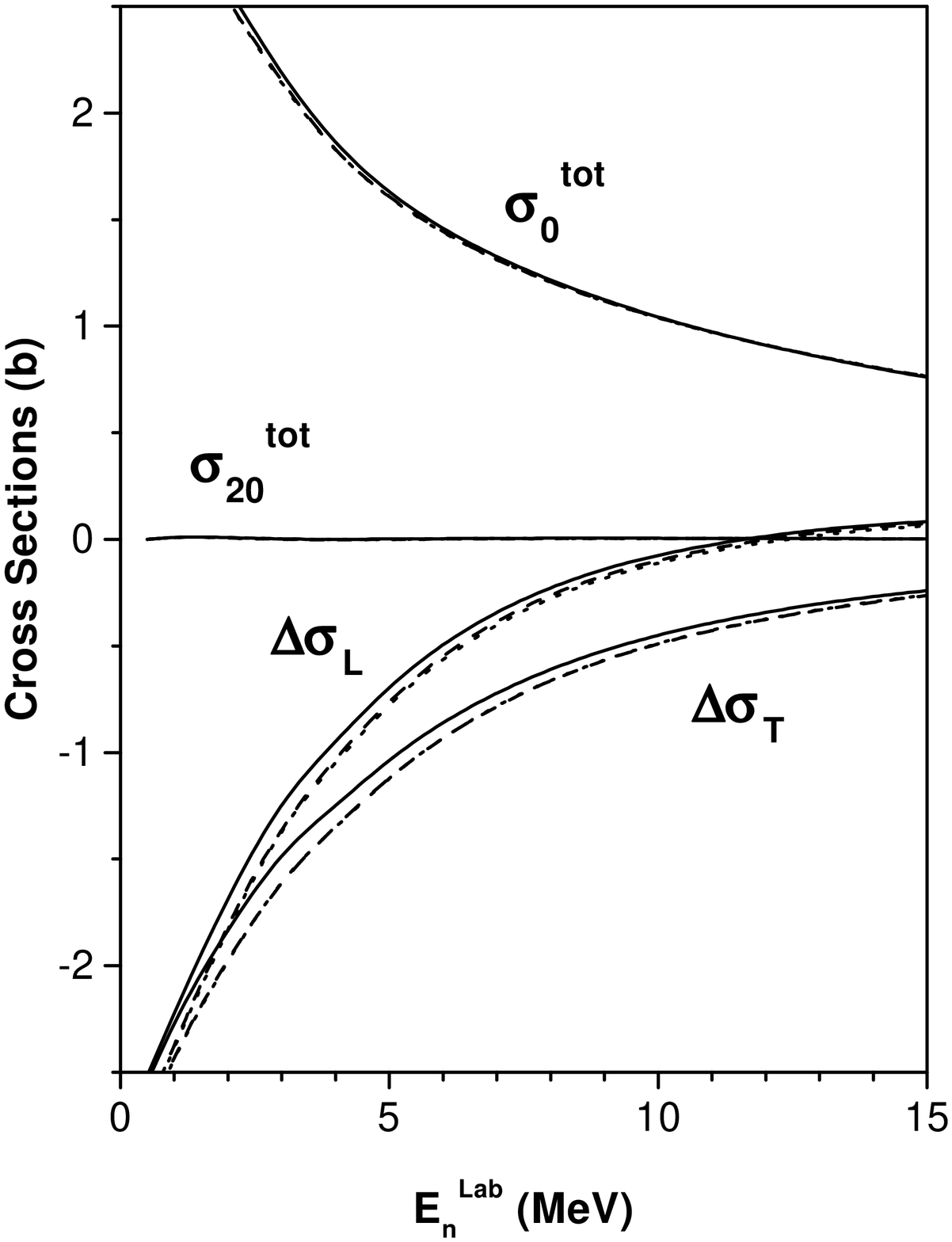,width=5in}
\caption{The total cross sections, $\sigma_0^{tot}$, $\Delta\sigma_L$, $\Delta\sigma_T$, and $\sigma_{20}^{tot}$, of the $nd$ scattering as a function of incident neutron energy in laboratory system for AV18 (solid lines), AV18+BR-3NF (dashed lines), and AV18+GS-3NF (dotted lines).
The dashed lines and the dotted ones are overlapped with each other.}
\label{fig:totcs}
\end{center}
\end{figure}

\begin{figure}
\begin{center}
\epsfig{file=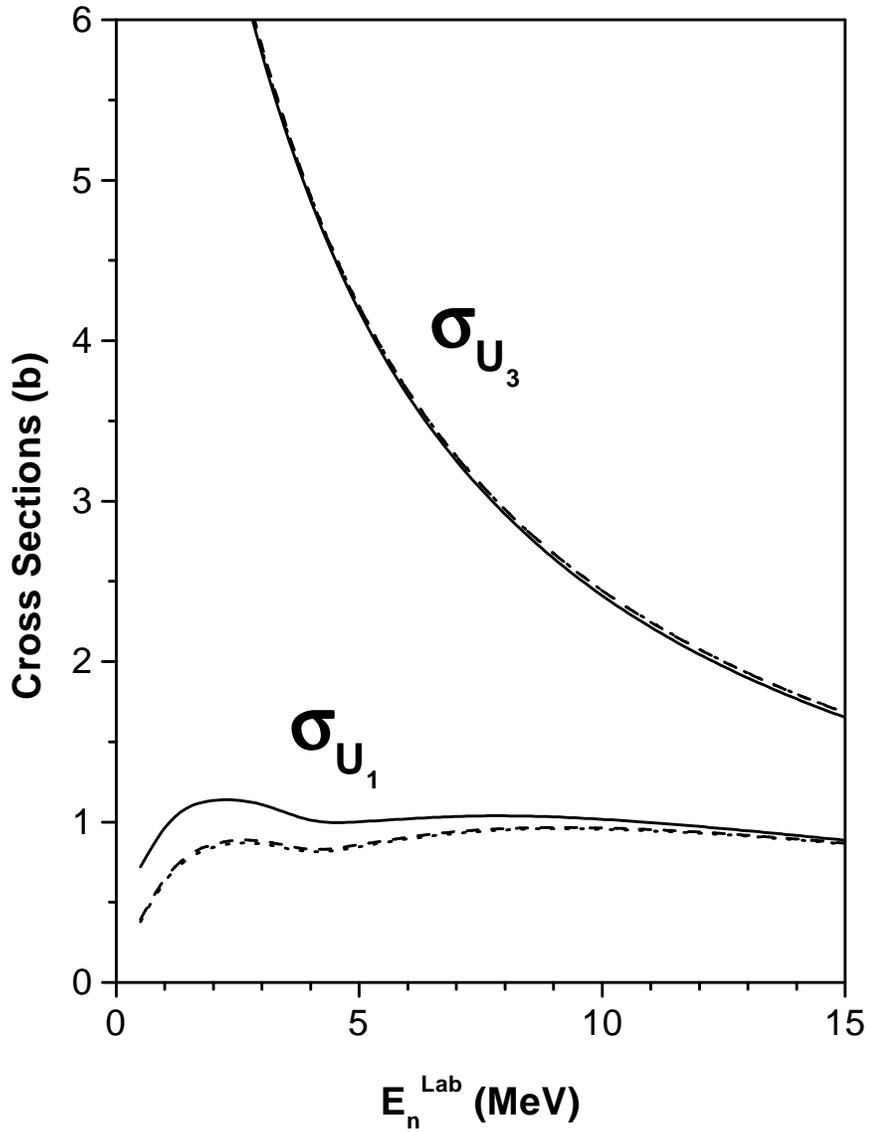,width=5in}
\caption{The total cross sections, $\sigma_{U_1}$, $\sigma_{U_3}$ of the $nd$ scattering as a function of incident neutron energy in laboratory system for AV18 (solid lines), AV18+BR-3NF (dashed lines), and AV18+GS-3NF (dotted lines).
The dashed lines and the dotted ones are overlapped with each other.}
\label{fig:sigS}
\end{center}
\end{figure}

\begin{figure}
\begin{center}
\epsfig{file=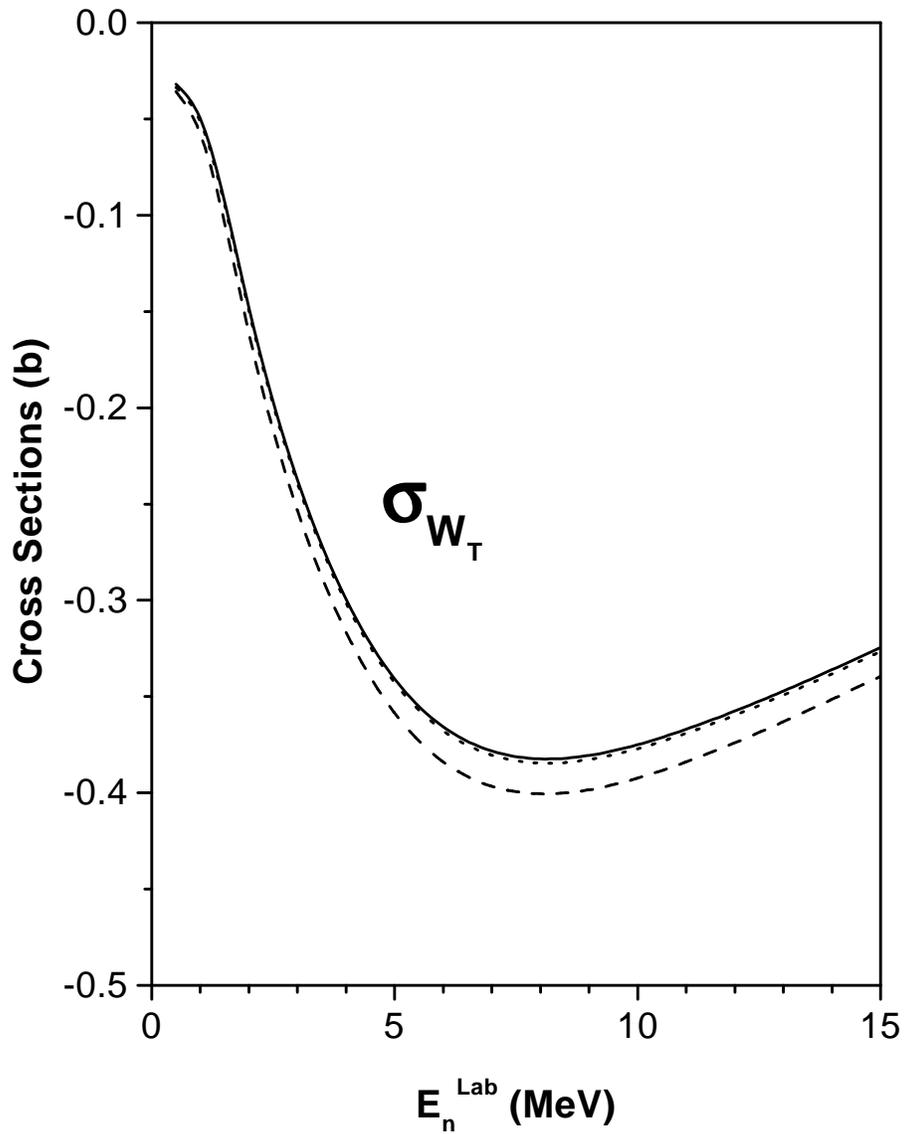,width=5in}
\caption{The total cross section, $\sigma_{W_T}$, of the $nd$ scattering as a function of incident neutron energy in laboratory system for AV18 (solid line), AV18+BR-3NF (dashed line), and AV18+GS-3NF (dotted line).}
\label{fig:sigW}
\end{center}
\end{figure}

In Fig.\ \ref{fig:totcs}, the calculated cross sections, $\sigma_0^{tot}$, $\Delta\sigma_L$, $\Delta\sigma_T$, and $\sigma_{20}^{tot}$, are shown as functions of the neutron incident energy up to $15$ MeV.
In the figure, one can see that the cross section $\sigma_{20}^{tot}$, which is expected to be sensitive to the deuteron $D$-state, is small compared to other three cross sections.
In Figs.\ \ref{fig:sigS} and \ref{fig:sigW}, ${\rm Im}(U_1)$, ${\rm Im}(U_3)$,  and ${\rm Im}(W_T)$ are shown by cross sections, $\sigma_A = \alpha {\rm Im}(A)$, where $A$ is  $U_1$ etc.
In Fig.\ \ref{fig:sigS}, the effect of 3NF on the scalar amplitude for the quartet state, $U_3$, is very small, while that for the doublet state, $U_1$, is remarkable, particularly at low incident energies.
In the latter, however, we cannot distinguish the effect of the BR-3NF from that of the GS-3NF.
This fact corresponds to the well known correlation between the calculated values of the triton binding energy and those of the $nd$ doublet scattering length \cite{Fr84,Is99}, which  means that the low energy doublet scattering is governed essentially by the position of the 3N bound state pole.

On the other hand, the effect of 3NF on the tensor amplitude $W_T$ has an interesting feature.
From Fig.\ \ref{fig:sigW}, one can see that the effect of the BR-3NF on $W_T$ is quite appreciable at large incident energies.
However that of the GS-3NF is almost negligible.
This is what to be expected from the consideration by the folding model, in which $W_T$ is directly related with the tensor interaction.

In summary, we have shown that the four non-vanishing independent forward amplitudes in the  $nd$ elastic scattering consist of two scalar amplitudes and two tensor amplitudes, which are respectively related to the two central interactions and the two tensor ones, and that one can determine experimentally the imaginary parts of these scattering amplitudes by measuring four cross sections, $\sigma_0^{tot}$, $\Delta\sigma_L$, $\Delta\sigma_T$, and $\sigma_{20}^{tot}$.
By the Faddeev calculation, the spin dependence of the 3NF contribution is investigated and it is found that one of the tensor amplitude, $W_T$, is most sensitive to the 3NF except for very low energies and it provides the information of the tensor effect of the 3NF.
These predictions will be encouraging the measurements of the total cross sections to obtain a significant information of the interaction between three nucleons.


\end{document}